\documentclass[11pt,twoside]{article}

\usepackage{asp2021}

\aspSuppressVolSlug
\resetcounters

\begin{document}

\title{Refactoring the SIXTE simulator: Towards a more modular code base}

\author{C.~Kirsch, M.~Lorenz, L.~Dauner, E.~G.~G\"ulbahar, K.~Pal, N.~Reinmann, J.~J.~R.~Stierhof, P.~Thalhammer, T.~Dauser and J.~Wilms}
\affil{Dr. Remeis-Sternwarte and ECAP, Friedrich-Alexander-Universit\"at Erlangen-N\"urnberg, Sternwartstr. 7, Bamberg, Bavaria, Germany; \email{christian.ck.kirsch@fau.de}}

\paperauthor{C.~Kirsch}{christian.ck.kirsch@fau.de}{0000-0001-6293-1538}{Friedrich-Alexander-Universit\"at Erlangen-N\"urnberg}{Dr. Remeis-Sternwarte and ECAP}{Bamberg}{Bavaria}{96049}{Bamberg}
\paperauthor{M.~Lorenz}{maximilian.ml.lorenz@fau.de}{0000-0001-6553-9590}{Friedrich-Alexander-Universit\"at Erlangen-N\"urnberg}{Dr. Remeis-Sternwarte and ECAP}{Bamberg}{Bavaria}{96049}{Bamberg}
\paperauthor{L.~Dauner}{lea.dauner@fau.de}{0000-0002-8735-9707}{Friedrich-Alexander-Universit\"at Erlangen-N\"urnberg}{Dr. Remeis-Sternwarte and ECAP}{Bamberg}{Bavaria}{96049}{Bamberg}
\paperauthor{E.~,G.~,G\"ulbahar}{esin.gulbahar@fau.de}{0009-0005-3584-1760}{Friedrich-Alexander-Universit\"at Erlangen-N\"urnberg}{Dr. Remeis-Sternwarte and ECAP}{Bamberg}{Bavaria}{96049}{Bamberg}
\paperauthor{K.~Pal}{karan.pal@fau.de}{0000-0002-6514-7033}{Friedrich-Alexander-Universit\"at Erlangen-N\"urnberg}{Dr. Remeis-Sternwarte and ECAP}{Bamberg}{Bavaria}{96049}{Bamberg}
\paperauthor{N.~Reinmann}{neo.reinmann@fau.de}{}{Friedrich-Alexander-Universit\"at Erlangen-N\"urnberg}{Dr. Remeis-Sternwarte and ECAP}{Bamberg}{Bavaria}{96049}{Bamberg}
\paperauthor{J.~J.~R.~Stierhof}{jakob.stierhof@fau.de}{0000-0001-9537-7887}{Friedrich-Alexander-Universit\"at Erlangen-N\"urnberg}{Dr. Remeis-Sternwarte and ECAP}{Bamberg}{Bavaria}{96049}{Bamberg}
\paperauthor{P.~Thalhammer}{philipp.thalhammer@fau.de}{0000-0001-6269-2821}{Friedrich-Alexander-Universit\"at Erlangen-N\"urnberg}{Dr. Remeis-Sternwarte and ECAP}{Bamberg}{Bavaria}{96049}{Bamberg}
\paperauthor{T.~Dauser}{thomas.dauser@fau.de}{0000-0003-4583-9048}{Friedrich-Alexander-Universit\"at Erlangen-N\"urnberg}{Dr. Remeis-Sternwarte and ECAP}{Bamberg}{Bavaria}{96049}{Bamberg}
\paperauthor{J.~Wilms}{joern.wilms@fau.de}{0000-0003-2065-5410}{Friedrich-Alexander-Universit\"at Erlangen-N\"urnberg}{Dr. Remeis-Sternwarte and ECAP}{Bamberg}{Bavaria}{96049}{Bamberg}

%\aindex{Kirsch,~C.}
%\aindex{Lorenz,~M.}
%\aindex{Dauner,~L.}
%\aindex{G\"ulbahar,~E.~G.}
%\aindex{Reinmann,~N.}
%\aindex{Pal,~K.}
%\aindex{Stierhof,~J.~J.~R.}
%\aindex{Thalhammer,~P.}
%\aindex{Dauser,~T.}
%\aindex{Wilms,~J.}

\begin{abstract}
The SIXTE (SImulation of X-ray TElescopes) software is a general end-to-end simulation toolkit for X-ray observations, covering the full observation process from source photon generation to detector readout and the production of high-level output files.
It is the official simulator for existing and future X-ray missions, such as \textit{eROSITA}, \textit{NewAthena}, \textit{THESEUS} and \textit{AXIS}.

Originally being designed as a simulator for \textit{eROSITA}, the addition of new instrument and telescope types over several years has made the original code base increasingly difficult to maintain.
As such, we have refactored the code, changing languages from C to C\texttt{++} and switching to a more modular software design to facilitate the implementation of new models.
This proceeding highlights some of the design choices used during the refactoring as well as its effects on maintenance and new feature development one year after release of the refactored code base.
\end{abstract}

%\ssindex{software!simulation}
%\ssindex{astronomy!X-ray}
%\ssindex{computer languages!C++}
%\ssindex{software!development}
%\ssindex{software!design}

%\ooindex{SIXTE, ascl:1903.002}

\section{Introduction}

The SIXTE \citep{2019A&A...630A..66D} software is an open source\footnote{\url{https://www.sternwarte.uni-erlangen.de/sixte/}} simulation toolkit for X-ray observations, with the goal of accurately simulating both existing and future X-ray missions in order to assess their scientific performance.
To achieve this, SIXTE generates source photons based on a flexible source definition format called SIMPUT.
Photons are then imaged onto the focal plane of a telescope using either raytracing or stored images of the telescope point spread function (PSF).
This imaging also takes into account effects due to off-axis pointings, e.g., during surveys, such as vignetting and PSF degradation.
The instrument models of SIXTE also take into account non-linear detection effects such as pile-up.
Figure \ref{fig:ero_crab} shows an example simulation with these effects.

  \articlefigurethree{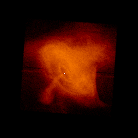}{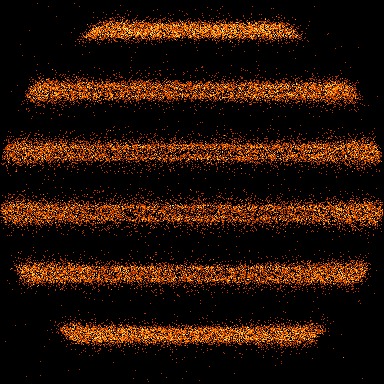}{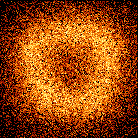}{fig:ero_crab}{Example simulation showing the capabilities of SIXTE for an extremely bright source.
  \emph{Left:} Input photons generated from a model containing the Crab Nebula and Pulsar.
  \emph{Middle:} Photons detected by one CCD of \textit{eROSITA} observing in survey mode, in raw detector coordinates.
  \emph{Right:} Detected photons reprojected into sky coordinates.
  The overall image is distorted by the optics of \textit{eROSITA}, with a loss of counts towards the center of the image due to pile-up.
  }

First versions of SIXTE were developed in the mid 2000's as separate tools for simulating X-ray observations.
Over the following roughly 20 years, the scope of the software was extended, starting with the simulation of CCD instruments like \textit{eROSITA} \citep{2021A&A...647A...1P}, and going to DEPFET detectors like the \textit{NewAthena Wide Field Imager (WFI)} \citep{2021SPIE11444E..0TM}, microcalorimeters like the \textit{NewAthena X-ray Integral Field Unit (X-IFU)} \citep{2025ExA....59...18P} and other detector types.

The majority of these new detector types and other features was implemented within the scope of academic theses (e.g., \citet{2012PhDT.......640S}) by more than twenty contributors, who were mostly astrophysicists.
Both the regular fluctuation of the developer base and development constraints by thesis and instrument deadlines led to several design decisions that, while effective in the short term, hampered the long-term development of the code base, especially with regards to adding new features.

With the intended addition of new missions such as the \textit{Advanced X-ray Imaging Satellite (AXIS)} \citep{2023SPIE12678E..1ER} and the \textit{Transient High-Energy Sky and Early Universe Surveyor (THESEUS)} \citep{2018AdSpR..62..191A}, we decided that the SIXTE code base needed to be refactored to be more extensible and maintainable.
The rest of this proceeding will discuss the process of this refactoring and its outcomes.

\section{Refactoring}
\subsection{Issues with the code base} \label{sec:issues}

One of the primary issues of the old SIXTE code base was code duplication and inefficient re-use of existing functionality.
A primary example of this was the existence of individual simulation tools for several instruments.
Aside from a general simulator for CCD-based instruments, called \texttt{runsixt}, additional simulators existed for \textit{eROSITA} (composed of seven individual telescopes with one CCD focal plane each), \textit{NewAthena WFI} (using a single telescope with a focal plane composed of four DEPFET detectors) and \textit{NewAthena X-IFU} (using a single telescope with a microcalorimeter detector).
While these tools differed in details such as the number of detectors or types of detectors simulated, they shared many other tasks, such as the generation of input photons, or reprojection of photons from detector to sky coordinates.

Some of these tasks, such as photon generation, were separated into their own functions and thus reusable, but many others, such as photon reprojection were not, due to slight differences in, e.g., detector geometries.
In addition, when implementing new simulation tools, much of the code used to initialize structures like those used for photon generation was copied from existing tools to new tools (with slight modifications, if necessary), including memory allocation and error handling.

In practice, this separation of the simulator into different tools made it difficult to add new features, as it often required editing multiple simulation tool chains with sometimes subtle differences that an individual developer was not aware of.
One of the main goals of the refactoring was the elimination of cases like these.

\subsection{Refactoring goals and process}
There were five main goals for the refactoring process:
\begin{enumerate}
  \item Code duplication -- examples of which were shown above -- should be reduced, to make the code more maintainable.
  \item The overall design of the code should be opened up for extensions.
This was mainly motivated by future missions with new detector types that would need to be supported.
\item The output files of the code should stay compatible with previous versions.
Users of SIXTE often process its output with custom-written scripts, which should not be affected by this refactoring.
\item There should be no regressions in performance, particularly in run time.
\item As SIXTE is the official simulator of several future X-ray missions, it must still be possible to compile and run the code during the lifetime of these missions, which may be more than 20 years into the future.

\end{enumerate}

The first decision made in the refactoring process was to port the code base from C to C\texttt{++}.
This choice essentially helps to fulfill all of the mentioned goals:
Code duplication can be reduced in C\texttt{++} by proper use of the \texttt{std} library, as well as using constructors and destructors for memory management.
Correct use of templates and \texttt{virtual} interfaces further reduces code duplication and also serve to make a code base more extensible.
In terms of performance, C and C\texttt{++} are generally equivalent, and more so dominated by algorithmic complexity.
Being an ISO standardized language, a C\texttt{++} project can also be expected to still compile during the lifetime of the missions SIXTE supports.

To ensure compatibility with the output of previous versions, we created a comprehensive testing setup.
These tests checked both the direct output event files of SIXTE as well as derived products used by scientists, such as spectra and images.
The tests were automatically run via GitLab continuous integration on every git commit, with merges to the main development branch only allowed when all tests passed.
While not done automatically, these tests could also be used to discover possible performance regressions.

To properly make use of C\texttt{++}, all developers attended dedicated workshops on C\texttt{++} software design.
Following these workshops, we then first created an outline object hierarchy which would encompass the main tasks of the old SIXTE version, such as photon generation, photon imaging and photon detection.
These objects are then used by a new tool called \texttt{sixtesim}, which replaces the individual simulator tools mentioned in Sec. \ref{sec:issues}.
To achieve this, \texttt{sixtesim} can flexibly simulate multiple telescopes at once (as needed for \textit{eROSITA}), which can each individually focus photons onto multiple detectors (as needed by \textit{NewAthena WFI}) of different types, such as CCD or DEPFET detectors and microcalorimeter arrays.
This was made possible by unifying the previously separate objects for simulating these different detector types behind an abstract \texttt{Detector} object using shared interfaces, but a separate internal implementation.

As a test of this hierarchy, we then implemented a full instrument setup -- in our case, a CCD detector based on \textit{eROSITA}.
This implementation helped diagnose any possible issues with the  object hierarchy before adding more detectors to it.

Once this first simulation chain was implemented, we could build out the object hierarchy by adding more detector types and develop further tools using this object hierarchy.
As the interfaces to all objects were defined ahead of time, this could be done in parallel by multiple developers.

\section{Summary and conclusions}
The implementation of the refactored SIXTE version took approximately two years, finishing in November 2024 with the public release of SIXTE version 3.0.3.
Development has continued since, with the latest version at the time of writing being version 3.3.2 in January 2026.

Aside from bug fixes, several new features were added in this time, mostly related to the \textit{THESEUS} mission, which uses both optics (lobster eye, coded mask) and detector types (silicon drift detectors) previously not supported by SIXTE.
For the optics, a new raytracing module was implemented, which can be used interchangeably with the old imaging implementation, based on the sampling of PSFs.

In terms of performance, the stated goal of preventing regressions was actually exceeded -- a review of the old code showed several inefficient algorithms and unnecessary memory allocations.
Improving these aspects of the code led to an improvement of the simulation speed by a factor of two to ten, depending on the used instrument and simulation case.
Testing during the refactoring also revealed several undiscovered bugs in the old SIXTE code, which were not reproduced in the new implementation.

In conclusion, the refactoring of the SIXTE code can be considered a success, providing a new code base that is much easier to maintain and extend, making it more robust for long term development and the inclusion of new X-ray instruments.

\bibliography{107}  % For BibTex

\end{document}